\documentclass{ws-ijmpa}

\begin{document}

\markboth{J. L. Albacete, N. Armesto, A. Capella, A. B. Kaidalov and C. A.
Salgado}
{Nuclear Shadowing and Diffraction}

%
\catchline{}{}{}{}{}
%

\title{NUCLEAR SHADOWING AND DIFFRACTION}

\author{J. L. ALBACETE}

\address{Departamento de F\'{\i}sica, M\'odulo C2, Planta baja,
Campus de Rabanales,\\ Universidad de C\'ordoba, 14071 C\'ordoba, Spain}

\author{\footnotesize \underline{N. ARMESTO} AND C. A. SALGADO}

\address{Department of Physics, CERN, Theory Division,
CH-1211 Gen\`eve 23, Switzerland}

\author{A. CAPELLA}

\address{Laboratoire de Physique Th\'eorique, Universit\'e de Paris XI,\\
B$\hat{a}$timent 210, F-91405 Orsay Cedex, France}

\author{A. B. KAIDALOV}

\address{Institute of Theoretical and Experimental Physics,\\ B.
Cheremushkinskaya
25, Moscow 117259, Russia}

\maketitle

\pub{Received (Day Month Year)}{Revised (Day Month Year)}

\begin{abstract}
The relation between diffraction in lepton-proton collisions and shadowing of
nuclear structure functions which arises from Gribov inelastic shadowing,
is described. A model realizing such relation, which produces a parameter-free
description of experimental data on nuclear structure functions at small $x$,
is presented.
The application to the description of multiplicities in nuclear collisions is
discussed and related to other approaches.

\end{abstract}

\section{Introduction and formalism}	
\label{intro}

The modification of parton densities inside nuclei
\cite{Arneodo:1992wf,Geesaman:1995yd} is one the most interesting
nuclear effects. It has strong practical implications on particle production
in nuclear collisions and also offers the
possibility to constrain theoretical models. We will be interested here
\cite{Armesto:2003fi} in the
region of small $x$, $x<0.01$, relevant for high energy collisions.
The observed feature in this region where isospin corrections are negligible
is that $F_{2A}<AF_{2p}$ i.e. the shadowing phenomenon.
Explanations of this shadowing range from leading twist to all twist effects.
Here (see \cite{Armesto:2003fi} for longer discussions, model description,
a more extensive
comparison with experimental data and other models, and full references)
we examine the possibility of describing
the small $x$ data on nuclear structure functions by the relation between
diffraction and nuclear shadowing which comes from inelastic shadowing
\cite{Gribov:1968jf,Gribov:1968gs,Abramovsky:1973fm}.

\begin{figure}[htb]
\centerline{\psfig{file=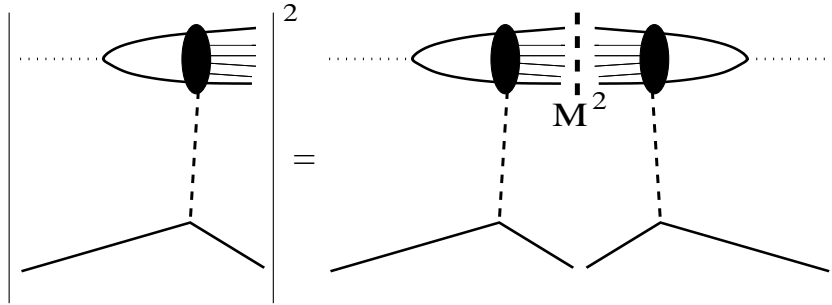,width=11cm,height=4cm}}
\vspace*{8pt}
\caption{Link between the diffractive cut and double scattering which produces
the first correction (nuclear shadowing)
to the additivity of cross section on a nucleus.}
\label{fig1}
\end{figure}

Reggeon Field Theory \cite{Gribov:1968fc} and its extension nuclei
\cite{Gribov:1968jf,Gribov:1968gs} are pre-QCD models in the form of field
theories, which have been very successful in describing experimental data on
hadronic collisions for the last 40 years.
In Fig.~\ref{fig1} we show the relation between the diagrams describing diffraction in
$\gamma^*$-nucleon collisions and double rescattering of the
$\gamma^*$ on a nucleus. In such relation the cut through an intermediate
state of the hadronic component of the $\gamma^*$
with arbitrary mass $M^2$ occurs, the so-called Gribov
inelastic shadowing \cite{Gribov:1968jf,Gribov:1968gs}. Through the AGK
cutting rules \cite{Abramovsky:1973fm} the contribution of the cut between the
amplitudes exchanged with the nucleus is minus the total 2-scattering
contribution,
\begin{equation}
\sigma_A^{(2)}= -4\pi A(A-1)\int d^2b \ T^2_A(b)
\int_{M_{min}^2}^{M_{max}^2} dM^2
\left.\frac{d\sigma^{\cal{D}}}{dM^2dt}
\right|_{t=0} F_A^2(t_{min}),
\label{eq1}
\end{equation}
with $M_{min}^2=4m_{\pi}^2=0.08$ GeV$^2$,
$M_{max}^2$ fixed by $x_P<0.1$, $T_A(b)$ the nuclear profile function
normalized to 1 and coherence effects, important for not so small $x>
(R_Am_N)^{-1}$, included in
\begin{equation}
F_A(t_{min})=\int d^2b J_0(b\sqrt{-t_{min}})T_A(b),\ \
t_{min}=-m_N^2 x_P^2.
\label{eq2}
\end{equation}
Higher order rescatterings are model-dependent \cite{Kopeliovich:2000ra}. To
estimate such uncertainty we use two models:
\begin{equation}
{\sigma_{\gamma^* A}^{Schw}(b) \over A\sigma_{\gamma^* N}}=
\frac{T_A(b)}{1+(A-1)f(x,Q^2)T_A(b)}> {\sigma_{\gamma^*
A}^{eik}(b)\over A\sigma_{\gamma^* N}}=
\frac{1-e^{-2(A-1)f(x,Q^2)T_A(b)}}{2(A-1)f(x,Q^2)}\, ,
\label{eq3}
\end{equation}
with $f(x,Q^2)$ chosen to match the two scattering result.

For the diffractive
cross section on the nucleon $d\sigma^{\cal{D}}/dM^2dt
|_{t=0}$, we use a model \cite{Capella:2000hq} which includes as
main ingredients a
separation between short distance (modeled in the dipole model) and long
distance (modeled by a pomeron + 1 lower Regge trajectory) contributions,
with separation
$r_0=0.2$ fm, and a
triple pomeron contribution required to describe diffraction. With 9 fitted
parameters it provides a good description of $F_2$ and $F_{2\cal{D}}$ for the
proton at
$x<0.01$, $Q^2<10$ GeV$^2$. The $t$-dependence of the diffractive cross section
is taken $\propto \exp{(Bt)}$ with $B=6$ GeV$^{-2}$.

\section{Nuclear structure functions}
\label{nsf}

\begin{figure}[htb]
\vskip -0.8cm
\centerline{\psfig{file=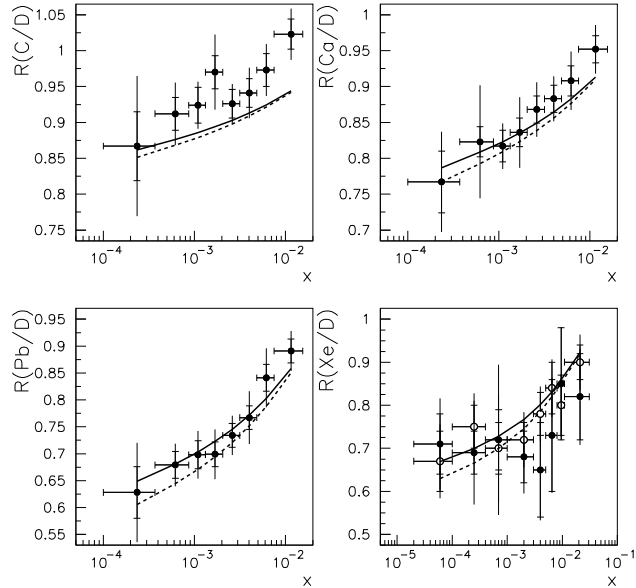,width=9cm}}
\vspace*{-8pt}
\caption{Results of the model using Schwimmer ($Schw$, solid lines) and
eikonal
($eik$, dashed lines) unitarization compared with experimental data versus $x$,
for the ratios
C/D, Ca/D,
Pb/D
and Xe/D
by the E665 Collaboration
(filled circles correspond to the
analysis with hadron requirement and open circles to that with
electromagnetic cuts, see the experimental paper for more details).}
\label{fig2}
\end{figure}

\begin{figure}[htb]
\centerline{\psfig{file=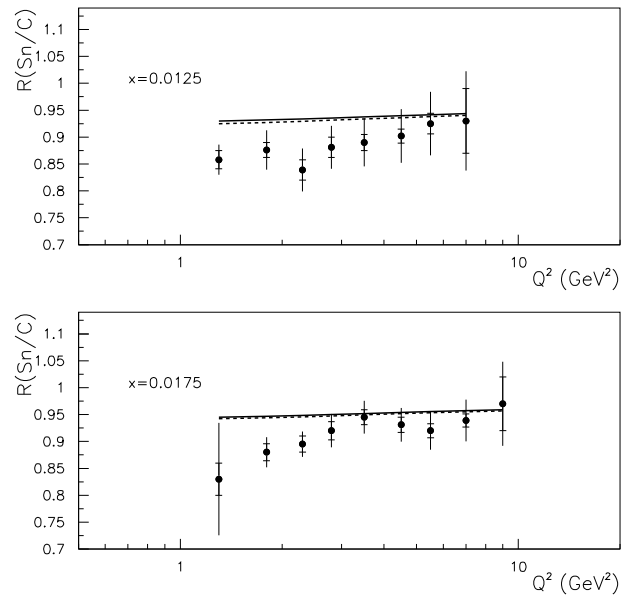,width=9cm}}
\vspace*{-8pt}
\caption{Results of the model using Schwimmer (solid lines) and
eikonal
(dashed lines) unitarization compared with experimental data versus $Q^2$,
for the ratio
Sn/C
by the NMC Collaboration
at two fixed small
values of $x$.}
\label{fig3}
\end{figure}

Within the explained formalism we obtain a parameter-free description of
shadowing in nuclear structure functions.
In Fig.~\ref{fig2} the results of the model for the ratios
$R(A/B)=(B\sigma_{\gamma^* A})/(A\sigma_{\gamma^* B})$
are compared with experimental
data for C/D, Ca/D,
Pb/D \cite{Adams:1995is} and Xe/D \cite{Adams:1992nf}. Both results, joined by
straight lines, and data have been obtained for a different $Q^2$ at each
value of $x$. Taking into account the absence of free parameters, the
agreement can be considered as quite good.
In Fig.~\ref{fig3} results of the
model are compared with data \cite{Arneodo:1996ru} on the $Q^2$ evolution of
the ratio Sn/C at fixed $x$. The model shows too flat a behavior, which may
point to the lack of perturbative (DGLAP) evolution to be eventually
applied for $Q^2$ of the order or larger than a few GeV$^2$.
Predictions for smaller $x$ which may be relevant
for future lepton-ion colliders can be found in \cite{Armesto:2003fi} together
with a comparison among available models,
which could be verified
through a measurement of $R(A/B)$
with $\sim 10$\% precision at $x\sim 10^{-4}$, $Q^2\sim 2$
GeV$^2$.

\section{Multiplicities}
\begin{figure}[htb]
\centerline{\psfig{file=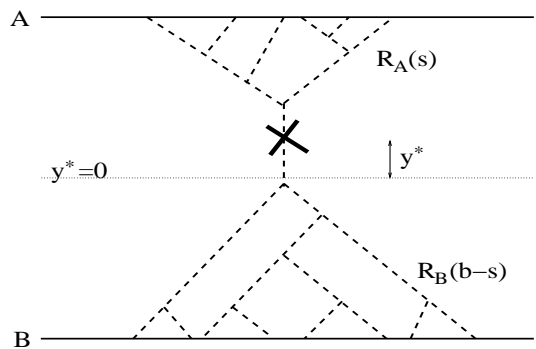,width=7cm,height=4.5cm}}
\vspace*{8pt}
\caption{Diagram showing the contribution to particle
production in AB collisions.}
\label{fig5}
\end{figure}

\begin{figure}[htb]
\centerline{\psfig{file=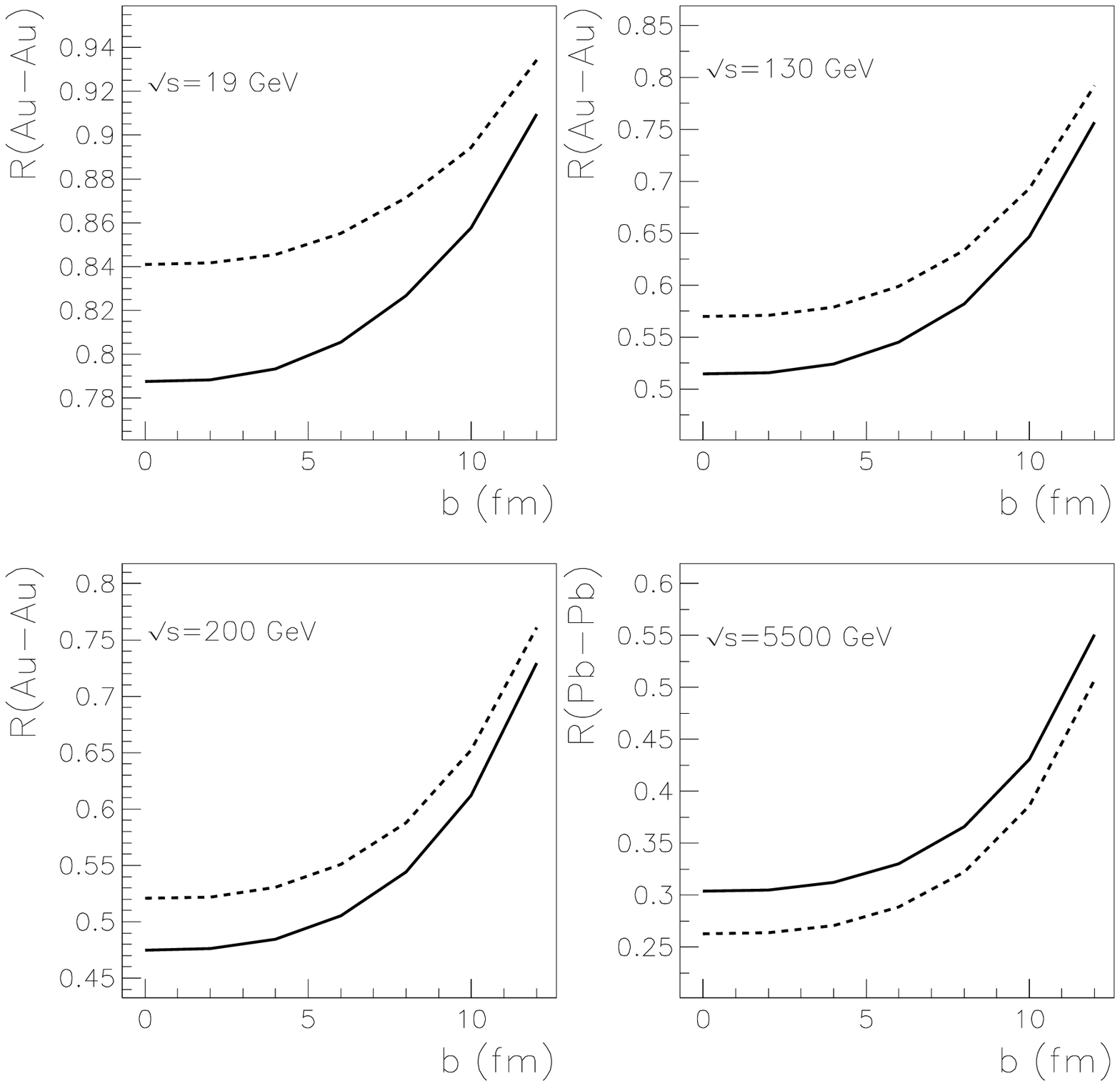,width=9cm}}
\vspace*{-8pt}
\caption{Results of the model for the multiplicity reduction factor
versus impact parameter $b$ at $y^*=0$, for AuAu collisions at $\sqrt{s}=19$,
130 and
200 GeV per nucleon, and for PbPb collisions at $\sqrt{s}=5500$ GeV
per nucleon,
in the parton model-like realization (solid lines) and for
a triple pomeron contribution alone
(dashed lines), see the text for explanations.}
\label{fig6}
\end{figure}

Using the AGK rules \cite{Abramovsky:1973fm}
it can be shown (Fig.~\ref{fig5})
within this kind of models that multiplicities are
proportional to the number of
nucleon-nucleon collisions $ABT_{AB}(b)/\sigma_{AB}(b)$ times a reduction
factor \cite{Caneschi:1975pg,Capella:1999kv}
\label{mult}
\begin{equation}
R_{AB}(b,y^*)=
\int d^2s \frac{R_A(\vec{s},y^*)R_B(\vec{s}-\vec{b},y^*)}{T_{AB}(b)},
\label{eq4}
\end{equation} 
with
$T_{AB}(b)=
\int d^2s \ T_A(\vec{s})T_B(\vec{s}-\vec{b})$,
$R_A(s,y^*)=T_A(s)/[1+(A-1)f(y^*,Q^2)T_A(s)]$, $R_A\propto A^{-1/3}$.
Function $f(y^*,Q^2)$ can be
computed in several ways \cite{Armesto:2003fi}: It can be taken from the
model described in the previous section but with integration limits inspired
by
the parton model: for projectile $A$ (target $B$),
$x_{A(B)}=m_T\,e^{\pm y^*}/\sqrt{s}$
with $y^*>0$ for the projectile hemisphere and $y^*<0$ for the target one,
$M_{max}^{2(A(B))}=Q^2\left(x_{Pmax}/x_{A(B)}-1\right)
=Q^2\left(x_{Pmax}\sqrt{s}\,e^{\mp y^*}/m_T-1\right)$,
$M^2_{min}$ fixed and equal to 0.08 GeV$^2$, and $Q^2=m_T^2=0.4$ GeV$^2$.
On the other hand, the reduction factor can be computed as in
\cite{Capella:1999kv}:
\begin{equation}
\label{eq6}
f(s,y^*)
=4\pi\int_{y_{min}}^{y_{max}}dy\
\frac{1}{\sigma_P(s)}\left.\frac{d\sigma^{PPP}}
{dydt}\right\vert_{t=0}F_A^2(t_{min}),
\end{equation}
with
$y^{(A(B))}_{min}$$=$$\ln{(s/M^{2(A(B))}_{max})}$
and $y_{max}^{(A(B))}$$=$$\ln{(s/m_T^2)}/2 \mp y^*$.
$\sigma_P(s)$ ($d\sigma^{PPP}/dydt$)
is the single (triple) Pomeron cross section
with parameters \cite{Armesto:2003fi} taken from \cite{Capella:1997yv}.
Fig.~\ref{fig6}
shows predictions for different symmetric nucleus-nucleus collisions
at $y^*=0$. The reduction factor
(\ref{eq4}) is needed
\cite{Capella:2000dn,alfons,Ferreiro:2004bq} to agree with experimental data.

\section{Comments}
\label{conclu}
In this contribution we have explored how lepton-nucleon
scattering can be related
to lepton-nucleus and nucleus-nucleus through
Gribov inelastic shadowing within the framework of Glauber-Gribov
theory.
Other relations have been found \cite{Armesto:2004ud}
within saturation physics. The model we have presented
shows some similarities to saturation ideas. Shadowing is taken into account
by similar diagrams. A factorization formula analogous to
(\ref{eq4}) has been shown
\cite{Kovchegov:2001sc} to hold in saturation for gluon production in
proton-nucleus collisions. It is also usually employed for nucleus-nucleus
e.g. in \cite{Kharzeev:2001gp},
but in this case corrections to factorization have been found
\cite{Krasnitz:1998ns,Balitsky:2004rr}. Studies at higher energies
on the centrality dependence of multiplicities at fixed and integrated
transverse momentum
will help to discriminate
between available approaches
\cite{Armesto:2000xh}.

\section*{Acknowledgments}
N.A. thanks
the organizers for their invitation to present this talk in such a
nice meeting.

\end{document}